# Transitioning water to an enhanced heat-conducting phase

James D. Brownridge

*Department of Physics, Applied Physics, and Astronomy, State University of New York at Binghamton, P. O. Box 6000, Binghamton, New York 13902-6000, USA*

Abstract

Water can be transitioned to an enhanced heat-conducting phase by supercooling only the water at the bottom of a container. The temperature gradient across the 4 cm in the center of an 8 cm long column of water with a 397 mW heat source at the top was lowered from 32°C to 0.75°C when the temperature at the bottom of the column was lowered from 1.2 °C to -5.6°C. The effective thermal conductivity of the water was increased from ~0.607 W/mK to ~24 W/mK. This result demonstrates that water has a high effective thermal conducting phase that has not been previously reported.

Many examples of methods that increase the effective thermal conductivity of water exist in the literature[1-7]. If the thermal conductivity of water is increased, then the amount of heat that can be transferred from one source to another in a given amount of time is also increased. This process is particularly important in water-cooled systems. Currently, the effective thermal conductivity of water is increased by adding impurities, usually nanoparticles[1,2,6]. Although the absolute thermal conductivity of water was not measured during the present study, an increase in the amount of heat conduction is evident in the data presented. For example, in Fig. 1 at time (b), the sudden decrease in the temperature of the water at the top of the column and the increase in the temperature at the bottom of the column, relative to the cooling bath's temperature, indicated that there was an increase in the flow of heat from the top of the water column to the cooling bath.

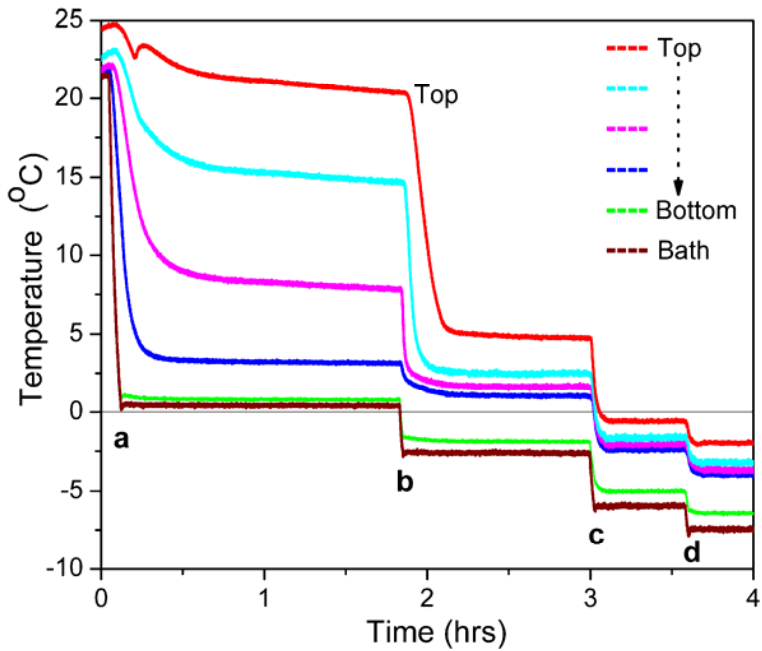

FIG. 1 At (a) the bottom tip of a column of water that was 8 cm tall (see Fig. 2) was confined in a glass test tube was inserted approximately 1 cm into a cooling bath that was held at 0.4°C. After remaining at 0.4°C for almost 2 hrs, the bath's temperature was lowered to -2.6°C at (b), and at (c), to -5.9°C and finally at (d), to -7.5°C. During this time, the top of the column of water was exposed to an ambient temperature of ~24°C. Note the abrupt drop in the temperature of the water at the top of the column when the water at the bottom of the column is transitioning into a supercool state by lowering only the temperature of the water at the bottom of the column to below 0°C at (b).

The data presented in Fig. 1 were acquired using a glass test tube that measured 10 cm in length and 1 cm in diameter. (See Fig. 2). This tube was filled with double-distilled water to yield a column of water that was 8 cm long. Type K thermocouples were located at 2-cm intervals from the bottom to the top of the test tube. A sixth thermocouple was positioned 2 mm below the bottom of the test tube to monitor the cooling bath's temperature relative to the bottom of the tube. The test tube was suspended above a variable-temperature bath, with the bottom of the test tube approximately 1 cm below the surface of the coolant. The bath temperature could be varied from -20$^o$C to 80$^o$C. Figure 1 illustrates the response of this system when the bottom tip of the test tube was inserted into the bath. The temperature of the cooling bath was stable at 0.4$^o$C. After approximately 6.5 min, the water at the bottom of the test tube had cooled to 0.7$^o$C (Fig. 1 (a)), and it remained at this temperature for the next two hours. It should be noted that the water in the bottom of the tube was approximately 0.3$^o$C warmer than that in the bath. This difference resulted from the flow of heat from the water in the tube to that in the bath. The ΔT between the water in the tube and the cooling bath increased as the heat flow into the bath increased. A change in the temperature of the water at the bottom of the column from 21.4$^o$C to 0.7$^o$C resulted in a change in the temperature at the top of the column of 4.2$^o$C over a two-hour period (Fig. 1 between (a) and (b)). After two hours, the temperature of the bath was lowered from 0.7$^o$C to -2.6$^o$C, and the temperature of the water at the top of the column decreased from 20.4$^o$C to 4.8$^o$C (Fig. 1 between (b) and (c)). This change was a 15.6$^o$C decrease in the temperature at the top of the tube and resulted from a net change in the water at the bottom of the tube of 2.5$^o$C. In addition, the ΔT between the bottom of the tube and the cooling bath increased to 0.84$^o$C from 0.30$^o$C, indicating an increase in the amount of heat flow from the tube to the cooling bath. The source of the heat was the ambient air because the top of the test tube was exposed to the air in the room. Additional decreases in the temperature of the water at the bottom of the column resulted in further decreases in the temperature at the top of the water column (Fig. 1 (c) to (d)). These decreases were not as dramatic as the transition from just above 0°C to just below 0°C.

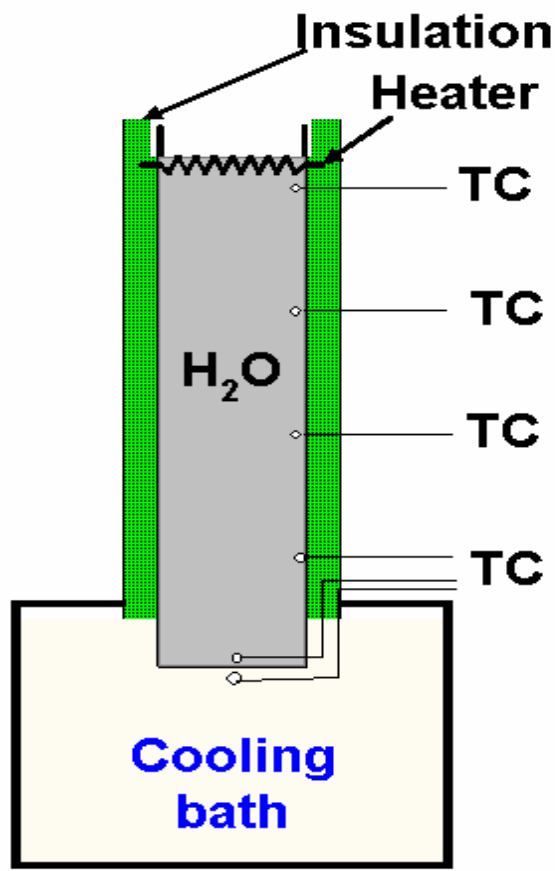

Fig. 2. Schematic illustration of a typical experimental set-up. The thermocouples (TC) are type K 0.005". The cooling bath is a Lauda/Brinkmann RM6 Refrigerating Circulating Bath.

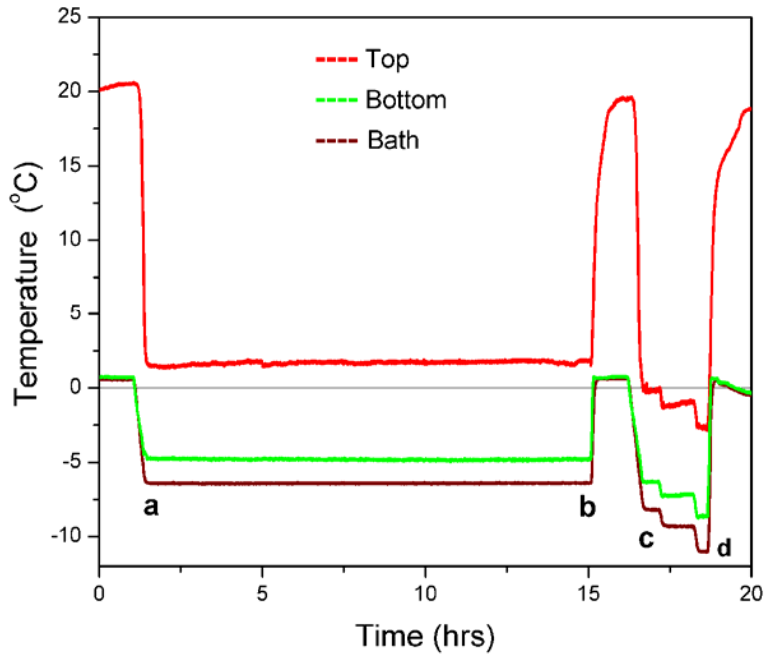

Fig. 3. At (a) water was switched into a high heat conduction state and remained there for about 14 hrs. The heat input was received from the ambient air at the top of the column of water. At (b) while the water temperature at the bottom of the column was increased from –4.7°C to 0.8°C, the water temperature at the top of the column increased from 1.7°C to 19.8°C. At (c) with the temperature of the entire water column below 0°C, a change in the water temperature at the bottom of the column of –2.3°C resulted in a change of only 2.4°C. The rate of heat conduction can only be changed by transitioning the temperature at the bottom of the column of water through 0°C, as shown at d.

If the temperature of the water at the bottom of the column was increased to just above 0°C, the temperature at the top of the column began to rise. When the bottom temperature was decreased to just below 0°C, the temperature at the top began to fall. However, if the water at the bottom began freezing, latent heat was released and caused a tremendous rise in the temperature of the water to 0°C. This well known phenomenon is demonstrated in Fig. 4 at (e). The release of latent heat induces the water out of the high heat conduction phase. In this case, the bottom of the column will freeze solid, and the top will approach ambient temperature and remain at this temperature until the ice at the bottom melts. When all of the ice is melted, the system resets, and a transition between the low and high heat conduction phases will occur as the bottom of the column transition up and down through 0°C.

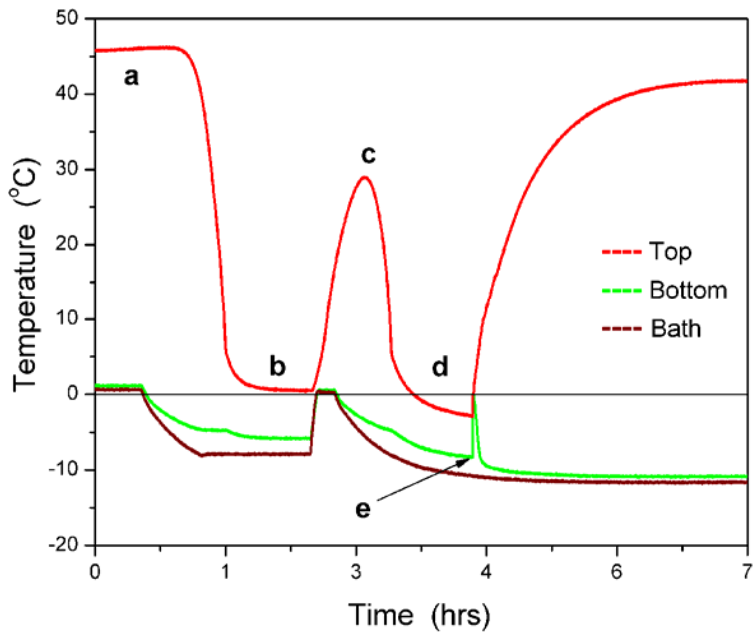

Fig. 4 The water was transition from a low (a, c) to a high (b, d) heat conduction state twice before latent heat was spontaneously released at −8.4°C (e). The latent heat temporarily increased the temperature of the water to 0°C, thereby transitioning the system out of the high heat conduction state. In this low heat conduction state, the water at the bottom of the column resumed cooling as the water at the top of the column increased in temperature toward 45°C. When all of the ice was melted, the water was returned to a high heat conduction state.

Water has many unique anomalies, but there are three that may be involved in the phenomenon that was observed in the present study[10,12]. These anomalies include that water's maximum density occurs at ~4°C, that water has the ability to supercool, and that water has the ability to have its heat capacity increased when it is supercool[13,14]. Supercooling is a phenomenon in which a material remains in a liquid state below its melting temperature. The present data suggest that when water transitions into a supercooled state, a change occurs that permits the coldest water clusters at the bottom of the column to move up the column at a much faster rate than they did before the transaction occurs. It is postulated that clusters[15,16,17] of water molecules with increased heat capacity[9,12] are able to move up the column more quickly than non-supercooled clusters that are just a few degrees warmer because these supercool clusters are expanded structures (ES) clusters and therefore less dense[15]. The density of an expanded structure is 0.94 g cm$^{-3}$ while a collapsed structure (CS) is 1.00 g cm$^{-3}$ [15]. A previous finding showed that when a few clusters of 99% pure heavy water are added to regular water, they inhibit or delay the movement of less dense water up the column[18], which tends to support this hypothesis. A similar, (inhibiting heat transfer) but reverse, process may have occurred in this case that involves non-supercooled and supercooled clusters of $H_2O$. However, the supercooled clusters are able to effectively transfer cold water up the column more freely than the non- supercooled clusters. In addition, these supercooled clusters carry colder water up the column more efficiently than the non-supercooled clusters. Experimental evidence also suggests[18] that heat is more efficiently transferred between these clusters when they are composed of the same type of water, i.e., homogeneous mixtures of HOH, HOD, DOH, and DOD vs. inhomogeneous mixtures of these types of water[19]. HOH is ordinary water also referred to as light water, DOD is heavy water, and here the two hydrogen atoms are replaced with two deuterium atoms. The HOD and DOH molecules each have single hydrogen and a single deuterium atom.

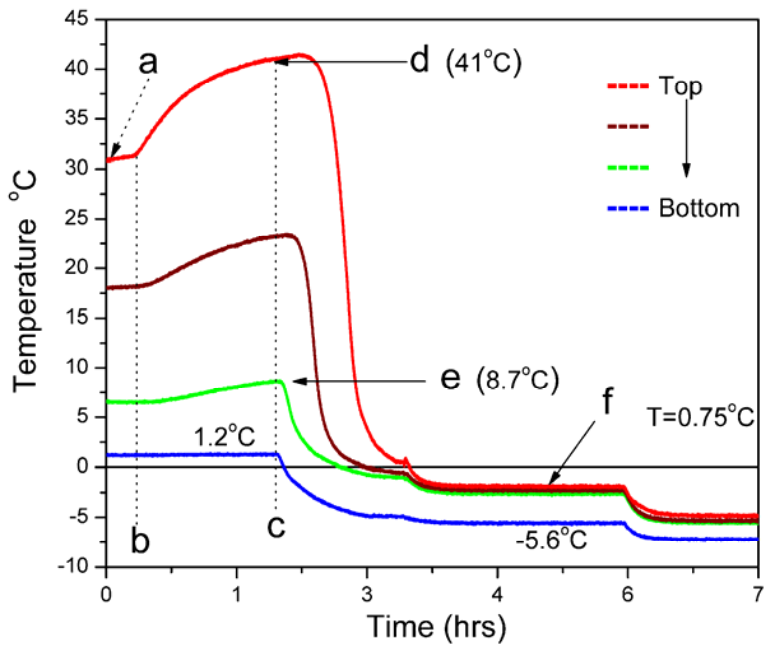

Fig. 5  The heat input at (a), was received from the ambient air at ~23°C plus 245 mW from a wire resistor heater that was wrapped around the outside top of the test tube. At (b), the power into the heater was increased to ~397 mW. At (c), the process of transitioning the water at the bottom of the column to a supercool state was initiated. At (f), the thermal gradient between (d), and (e), had decreased from 32.3°C to a gradient of only 0.75°C over a water column length of 4 cm.

To test this hypothesis, HOH was replaced with DOD, and a similar phenomenon was observed when the heavy water was supercooled at the bottom of the column. However, when light water was carefully placed on top of the heavy water to produce a 2-cm-tall column of light water on top of a 6-cm-tall column of heavy water, the heavy water cluster was unable to penetrate the boundary between the two types of water. The boundary consisted of DOD, HOD, DOH, and HOH. However, after the two types of water were mixed by vigorous shaking, they produced a homogeneous mixture. As a result, the supercooled clusters (all of the clusters were the same type of water at this point) that formed at the bottom of the column were able to move to the top of the column. This finding was further tested with a heat sources at the top of a column of light water. An example is shown in Fig. 5, which illustrates how the system responded when the input heat energy was changed from ~245 mW to ~397 mW and when the water was transition into its high heat conduction phase. The thermal gradient across the center 4 cm of the 8-cm column of water decreased from ~32.3$^o$C to <0.75$^o$C, which is greater than a 99% decrease. Supercooling water at the bottom of a column of water can effectively reduce the water's resistance to heat flowing through it from the top to the bottom and can lower this resistance to almost zero in the column center.

**Acknowledgements** The author thanks Profs. B. White and S. Shafroth for many discussions and suggestions. This work was supported by Binghamton University's Department of Physics, Applied Physics, and Astronomy.

jdbjdb@binghamton.edu

References

1. J. A. Eastman, S. U. S. Choi, S. Li, W. Yu, and L. J. Thompson, Appl. Phys. Lett. **78**, 718-720 (2009).

2. S. A. Putman, D. G. Cahill, and P. V. Braun, J. Appl. Phys. **99**, 084308-6 (2006).

3. C. H. Chon, K. D. Kihm, S. P. Lee, and S. U. S. Choi, Appl. Phys. Lett. **87**, 153107-3 (2005.

4. Y. Xuan and Q. Li. Int. J. Heat and Fluid Flow. **21**, 58-57 (2000).

5. R. Prasher, P. Bhattavharya, P. Phelan, Phys. Rev. Lett. **94**, 025901-4 (2006).

6. H. Xie, H. Lee, W. Youn and M. Choi, J. Appl. Phys. **94**. 4967-5 (2003).

7. B. Yang, and Z. H. Han, Appl. Phys. Lett. **88**, 261914-3 (2006).

8. M. F. Chaplin, www.lsbu.ac.uk/water/data.html. (2011).

9. R. Ludwig, Angew. Chem. Int. **40**, 1808-1827 (2001).

10. M. F. Chaplin, www.lsbu.ac.uk/water/anmlies.html. (2011).

11. O. Mishima and H. F. Stanley, Nature **396**, 330-335 (1998).

12. C. A. Angell, J. Shuppert, and J. C. J. Tucker, phys. Chem. **77**, 3092-3099 (1973).

13. R. Bergman and J. Swenson, Nature **403**, 283-286 (2000).

14. E. Tombari, C. Ferrari, and G. Salvetti, Chem. Phys. Lett. **300**, 749-751 (1999).

15. M. F. Chaplin, A proposal for the structure of water. Biophysical Chem. **83**, 211-221 (1999).

17. F. Sciortino, A. Giger, and H. E. Stanley, Nature **354**, 218-221 (1991).

18. W. R. Gorman and J. D. Brownridge, Appl. Phys. Lett. 93, 034101-3 (2008).

19. A.K. Soper and C. J. Benmore, Phys, Rev, Lett. **101**, 065502-4 (2008).